\documentclass[aps,pra,reprint,superscriptaddress,nofootinbib,longbibliography,floatfix]{revtex4-2}

\usepackage{amsmath}
\usepackage{graphicx}
\usepackage{color}
\usepackage[10pt]{moresize}
\usepackage{amsfonts}
\usepackage{amssymb}
\usepackage{amscd}
\usepackage{enumerate}
\usepackage{epsfig}
\usepackage{subfigure}
\usepackage{graphicx}
\usepackage{bm}
\usepackage{soul}
\usepackage{ulem}
\usepackage{multirow}
\graphicspath{{../}}

\definecolor{BLUE}{rgb}{0,0,1}

\begin{document}

\title{High fidelity control of superconducting qubits with optical transmitted signal}

\author{Yu-Huai Li}
\thanks{These authors contributed equally to this work}
\affiliation{Hefei National Research Center for Physical Sciences at the Microscale and School of Physical Sciences, University of Science and Technology of China, Hefei 230026, China.}
\affiliation{Shanghai Research Center for Quantum Science and CAS Center for Excellence in Quantum Information and Quantum Physics, University of Science and Technology of China, Shanghai 201315, China.}
\affiliation{Hefei National Laboratory, University of Science and Technology of China, Hefei 230088, China.}

\author{Daojin Fan}
\thanks{These authors contributed equally to this work}
\affiliation{Hefei National Research Center for Physical Sciences at the Microscale and School of Physical Sciences, University of Science and Technology of China, Hefei 230026, China.}
\affiliation{Shanghai Research Center for Quantum Science and CAS Center for Excellence in Quantum Information and Quantum Physics, University of Science and Technology of China, Shanghai 201315, China.}
\affiliation{Hefei National Laboratory, University of Science and Technology of China, Hefei 230088, China.}

\author{Na Li}
\affiliation{Hefei National Research Center for Physical Sciences at the Microscale and School of Physical Sciences, University of Science and Technology of China, Hefei 230026, China.}
\affiliation{Shanghai Research Center for Quantum Science and CAS Center for Excellence in Quantum Information and Quantum Physics, University of Science and Technology of China, Shanghai 201315, China.}
\affiliation{Hefei National Laboratory, University of Science and Technology of China, Hefei 230088, China.}

\author{Fusheng Chen}
\affiliation{Hefei National Research Center for Physical Sciences at the Microscale and School of Physical Sciences, University of Science and Technology of China, Hefei 230026, China.}
\affiliation{Shanghai Research Center for Quantum Science and CAS Center for Excellence in Quantum Information and Quantum Physics, University of Science and Technology of China, Shanghai 201315, China.}
\affiliation{Hefei National Laboratory, University of Science and Technology of China, Hefei 230088, China.}

\author{Shaowei Li}
\affiliation{Hefei National Research Center for Physical Sciences at the Microscale and School of Physical Sciences, University of Science and Technology of China, Hefei 230026, China.}
\affiliation{Shanghai Research Center for Quantum Science and CAS Center for Excellence in Quantum Information and Quantum Physics, University of Science and Technology of China, Shanghai 201315, China.}
\affiliation{Hefei National Laboratory, University of Science and Technology of China, Hefei 230088, China.}

\author{Dong-Dong Li}
\affiliation{Hefei National Research Center for Physical Sciences at the Microscale and School of Physical Sciences, University of Science and Technology of China, Hefei 230026, China.}
\affiliation{Shanghai Research Center for Quantum Science and CAS Center for Excellence in Quantum Information and Quantum Physics, University of Science and Technology of China, Shanghai 201315, China.}
\affiliation{Hefei National Laboratory, University of Science and Technology of China, Hefei 230088, China.}
\affiliation{QuantumCTek Co., Ltd., Hefei 230088, China.}

\author{Yu Xu}
\affiliation{Hefei National Research Center for Physical Sciences at the Microscale and School of Physical Sciences, University of Science and Technology of China, Hefei 230026, China.}
\affiliation{Shanghai Research Center for Quantum Science and CAS Center for Excellence in Quantum Information and Quantum Physics, University of Science and Technology of China, Shanghai 201315, China.}
\affiliation{Hefei National Laboratory, University of Science and Technology of China, Hefei 230088, China.}

\author{Jin Lin}
\affiliation{Hefei National Research Center for Physical Sciences at the Microscale and School of Physical Sciences, University of Science and Technology of China, Hefei 230026, China.}
\affiliation{Shanghai Research Center for Quantum Science and CAS Center for Excellence in Quantum Information and Quantum Physics, University of Science and Technology of China, Shanghai 201315, China.}
\affiliation{Hefei National Laboratory, University of Science and Technology of China, Hefei 230088, China.}

\author{Ming Gong}
\affiliation{Hefei National Research Center for Physical Sciences at the Microscale and School of Physical Sciences, University of Science and Technology of China, Hefei 230026, China.}
\affiliation{Shanghai Research Center for Quantum Science and CAS Center for Excellence in Quantum Information and Quantum Physics, University of Science and Technology of China, Shanghai 201315, China.}
\affiliation{Hefei National Laboratory, University of Science and Technology of China, Hefei 230088, China.}

\author{He-Liang Huang}
\affiliation{Henan Key Laboratory of Quantum Information and Cryptography, Zhengzhou, Henan 450000, China}

\author{Hui Deng}
\affiliation{Hefei National Research Center for Physical Sciences at the Microscale and School of Physical Sciences, University of Science and Technology of China, Hefei 230026, China.}
\affiliation{Shanghai Research Center for Quantum Science and CAS Center for Excellence in Quantum Information and Quantum Physics, University of Science and Technology of China, Shanghai 201315, China.}
\affiliation{Hefei National Laboratory, University of Science and Technology of China, Hefei 230088, China.}

\author{Yulin Wu}
\affiliation{Hefei National Research Center for Physical Sciences at the Microscale and School of Physical Sciences, University of Science and Technology of China, Hefei 230026, China.}
\affiliation{Shanghai Research Center for Quantum Science and CAS Center for Excellence in Quantum Information and Quantum Physics, University of Science and Technology of China, Shanghai 201315, China.}
\affiliation{Hefei National Laboratory, University of Science and Technology of China, Hefei 230088, China.}

\author{Haoran Qian}
\affiliation{Hefei National Research Center for Physical Sciences at the Microscale and School of Physical Sciences, University of Science and Technology of China, Hefei 230026, China.}
\affiliation{Shanghai Research Center for Quantum Science and CAS Center for Excellence in Quantum Information and Quantum Physics, University of Science and Technology of China, Shanghai 201315, China.}
\affiliation{Hefei National Laboratory, University of Science and Technology of China, Hefei 230088, China.}

\author{Shaojun Guo}
\affiliation{Hefei National Research Center for Physical Sciences at the Microscale and School of Physical Sciences, University of Science and Technology of China, Hefei 230026, China.}
\affiliation{Shanghai Research Center for Quantum Science and CAS Center for Excellence in Quantum Information and Quantum Physics, University of Science and Technology of China, Shanghai 201315, China.}
\affiliation{Hefei National Laboratory, University of Science and Technology of China, Hefei 230088, China.}

\author{Futian Liang}
\affiliation{Hefei National Research Center for Physical Sciences at the Microscale and School of Physical Sciences, University of Science and Technology of China, Hefei 230026, China.}
\affiliation{Shanghai Research Center for Quantum Science and CAS Center for Excellence in Quantum Information and Quantum Physics, University of Science and Technology of China, Shanghai 201315, China.}
\affiliation{Hefei National Laboratory, University of Science and Technology of China, Hefei 230088, China.}

\author{Xiaobo Zhu}
\affiliation{Hefei National Research Center for Physical Sciences at the Microscale and School of Physical Sciences, University of Science and Technology of China, Hefei 230026, China.}
\affiliation{Shanghai Research Center for Quantum Science and CAS Center for Excellence in Quantum Information and Quantum Physics, University of Science and Technology of China, Shanghai 201315, China.}
\affiliation{Hefei National Laboratory, University of Science and Technology of China, Hefei 230088, China.}

\author{Cheng-Zhi Peng}
\affiliation{Hefei National Research Center for Physical Sciences at the Microscale and School of Physical Sciences, University of Science and Technology of China, Hefei 230026, China.}
\affiliation{Shanghai Research Center for Quantum Science and CAS Center for Excellence in Quantum Information and Quantum Physics, University of Science and Technology of China, Shanghai 201315, China.}
\affiliation{Hefei National Laboratory, University of Science and Technology of China, Hefei 230088, China.}

\author{Jian-Wei Pan}
\affiliation{Hefei National Research Center for Physical Sciences at the Microscale and School of Physical Sciences, University of Science and Technology of China, Hefei 230026, China.}
\affiliation{Shanghai Research Center for Quantum Science and CAS Center for Excellence in Quantum Information and Quantum Physics, University of Science and Technology of China, Shanghai 201315, China.}
\affiliation{Hefei National Laboratory, University of Science and Technology of China, Hefei 230088, China.}

\date{\today}

\begin{abstract}
Superconducting circuits exhibit remarkable potential for constructing large-scale quantum simulation and computation systems, featuring numerous qubits, extended coherence time, and precise control. Nevertheless, the growing number of signal cables poses a challenge in dilution refrigerators due to space and heat load constraints. To overcome this issue, we experimentally implemented an optically-assisted transmission line as an alternative to coaxial cables. By modulating microwave signals on laser intensities at room temperature and regenerating the signals at a cryogenic plate within the dilution refrigerator, we demonstrated full control of superconducting qubits using photocurrent. We demonstrate and benchmark both single-qubit and two-qubit gates on frequency tunable transmon qubits, 
achieving fidelities of 99.915\% \(\pm\) 0.005\% and 99.676\% \(\pm\) 0.041\%, respectively, which have reached the requirement of the surface code.
\end{abstract}

\maketitle

\section{Introduction}

Over the last several decades, tremendous progress has been made in building quantum computers \cite{Arute2019, science.abe8770, PhysRevLett.127.180501, Bernien2017, Zhang2017, Song2019, PhysRevLett.123.050502, PhysRevLett.122.110501, PhysRevLett.123.250503, PhysRevLett.120.260502, PhysRevLett.117.210502, Omran2019,2024Realization}.
In particular, the quantum processor architecture based on superconducting qubits has emerged as one of the leading candidates for scalable quantum computing platform \cite{ac-031119-050605, Wendin_2017, huang2020superconducting}.
The breakthrough of quantum computational advantage has been demonstrated in recent years \cite{Arute2019, science.abe8770, PhysRevLett.127.180501}.
In the next stage, along with the rapid development of noisy intermediate scale quantum (NISQ) technology \cite{Preskill2018quantumcomputingin} and to demonstrate the logic qubit through surface code error correction \cite{PhysRevA.86.032324, SurfaceCode22, Erhard2021, Andersen2020, Marques2022, Chen2021}, engineering a complete architecture for thousands or even more qubit scale superconducting quantum computing systems will become a critical and also tricky task.
Especially for a certain scale of superconducting qubit systems, a threshold of fidelity for one-qubit and two-qubit gates is required to realize a logic qubit with surface code.
To initialize the systems in the ground state and avoid thermal excitation errors during operations, the superconducting quantum computing chip needs to be installed inside a dilution refrigerator (DR) to achieve the temperature at the millikelvin level.
However, when scaling to large-scale quantum processors, the increasing number of microwave cables will pose a huge challenge to the heat load and space of DRs \cite{Krinner2019}.

Cryogenic electronics \cite{McDermott_2018, Leonard2019, bardin2019design, Xue2021} can be one of the possible approaches for the further scaling of superconducting quantum systems.
With the technical of cryo-CMOS or single flux quantum (SFQ), the control and readout signal of the qubit can be generated at low temperatures, thus dramatically reducing the wiring requirements from room-temperature devices.
However, many fundamental and engineering-related issues still need to be addressed for this technology.
Another approach is transmitting microwave control signals by photonics.
Benefiting from broadband and low transmission loss, visible and near-infrared (NIR) photons have become a widely used carrier for the generation, manipulation, and distribution of microwave signals.
In recent years, the technical of controlling microwaves via photonics has been rapidly developed and used in many applications, such as photonic true-time delay beamforming \cite{beamforming}, radio-over-fiber system \cite{RoF}, and photonic analog-to-digital conversion \cite{Valley:07}.
By installing photodiodes (PDs) in the cryogenic stage of the DR, the microwave signal can be directly transmitted through optical fiber and regenerated on the PDs without metallic coaxial cables.
With extremely low passive heat load, fibers are suitable for high-density wiring from room-temperature devices to quantum devices at cryogenic temperatures.
Efforts have been taken to demonstrate the possibility of optically carried microwave signals for superconducting qubits \cite{Lecocq2021, Youssefi2021, Delaney2022, li_optical_2024,0All2025,van_thiel_optical_2025,warner_coherent_2025,xie_scalable_2025}.

Here, we experimentally demonstrated the full controlling of frequency tunable transmon qubits \cite{barends2014superconducting,koch2007charge,gong2019genuine} by optically transmitted signals.
Installing the PDs, which regenerate microwave signals, at different stages of the DR was considered and compared according to the active heat load of the photonics link.
As a result, the most feasible configuration for a near-term application with thousands of qubits would be regenerating the control signal at the $4~K$ stage.
The fidelity of the single-qubit gate and two-qubit CZ gate is estimated as 99.915\% \(\pm\) 0.005\% and 99.676\% \(\pm\) 0.041\%, respectively, which has reached the threshold of surface code \cite{PhysRevA.86.032324,barends2014superconducting}.
With all the XY and Z control signals transmitted by optics, most of the metallic cables in traditional configuration can be replaced, thus tremendously increasing the capacity of the DR.

\section{Thermal-load considerations}

The primary source of the heat load at the cryogenic stage for such an RF photonic link is usually the active heat load, which is highly dependent on the signal's amplitude and duty cycle, and also scales with the number of qubits.
Waiting for an appropriately long time between experimental trials would be a possible approach to reduce the average active heat load \cite{Lecocq2021}, with the cost of lowering the repetition rate.
However, with the development of qubit fabrication and error correction, the lifetimes of the qubits are expected to be sufficiently long.
Thus, the heating effect would have already been significant enough during a single trial.
Besides, with the development of technologies such as active reset \cite{riste2012initialization,marques2023all,gao2025establishing}, removing the waiting time between trails has become a trend.
Therefore, it is imperative to consider the active heat load under the premise of a full-time running situation.
Typically, superconducting qubits require a peak power of $-66~dBm$ \cite{Krinner2019} (around $3~\mu A$) for the XY control signal, and up to hundreds of microamp for the fast bias for the tuning of interaction between qubits.
Thus, the active heat load of PDs for generating the above signals can range from a few to hundreds of microwatts for each qubit, which should be carefully considered. Further details are provided in Appendix~\ref{app:thermal}.
We take Bluefors XLD400 DR \cite{Krinner2019} as an example to consider the heat load.
A comparison of installing PDs at different stages of the DR was made, as shown in Tab. \ref{tab:1}.
The supported number of the optical channel, i.e., the number of qubits, is quantified by $P_C/P_L$, where $P_C$ is the cooling power and $P_L$ is the heat dissipation for each channel.
4 K and Still stages have the capacity to support the order of thousands of optical-assisted transmission lines and thus are good candidates.
It is worth noting that installing cryogenic electronics devices in the mixing chamber is also challenging due to the inevitable heat loads and other effects such as quasiparticle poisoning of the qubit \cite{Leonard2019}.
Thus, transmitting control signals from the 4 K or Still stage to the superconducting qubits is a common challenge faced by cryogenic electronics and optical transmission schemes.
Several mature superconducting cables, such as NbTi cables, have extremely low thermal conductors at this temperature region; thus, the potential can be further tapped.

\begin{table}[!tbp]\center
 \begin{tabular}{|c|c|c|c|c|c|}
 \hline
 \multirow{2}*{Stage} & \multirow{2}*{$P_C$ (mW)} & \multicolumn{2}{|c|}{XY only} & \multicolumn{2}{|c|}{XY and Z} \\
 \cline{3-6}
  & & $P_L$ (mW) & $P_C/P_L$ & $P_L$ (mW) & $P_C/P_L$ \\
 \hline
 4 K & 1,500 (at 4.2 K) & 0.058 & 25816 & 0.752 & 1996 \\
 Still & 40 (at 1.2 K) & 0.058 & 688 & 0.752 & 53.2 \\
 CP & 0.2 (at 140 mK) & 0.018 & 10.9 & 0.246 & 0.81 \\
 MXC & 0.019 (at 20 mK) & 0.006 & 3.3 & 0.073 & 0.24 \\
 \hline
 \end{tabular}
 \caption{Comparison of the cooling power $P_C$ of Bluefors XLD400 DR \cite{Krinner2019}, heat dissipation of each PD $P_L$ and the supported number of channel $P_C/P_L$ at different stages. The detailed calculation is provided in Appendix~\ref{app:thermal}.}
 \label{tab:1}
 \end{table}

\section{Optical transmission and cryogenic setup}

\begin{figure}[t]\center
\resizebox{8cm}{!}{\includegraphics{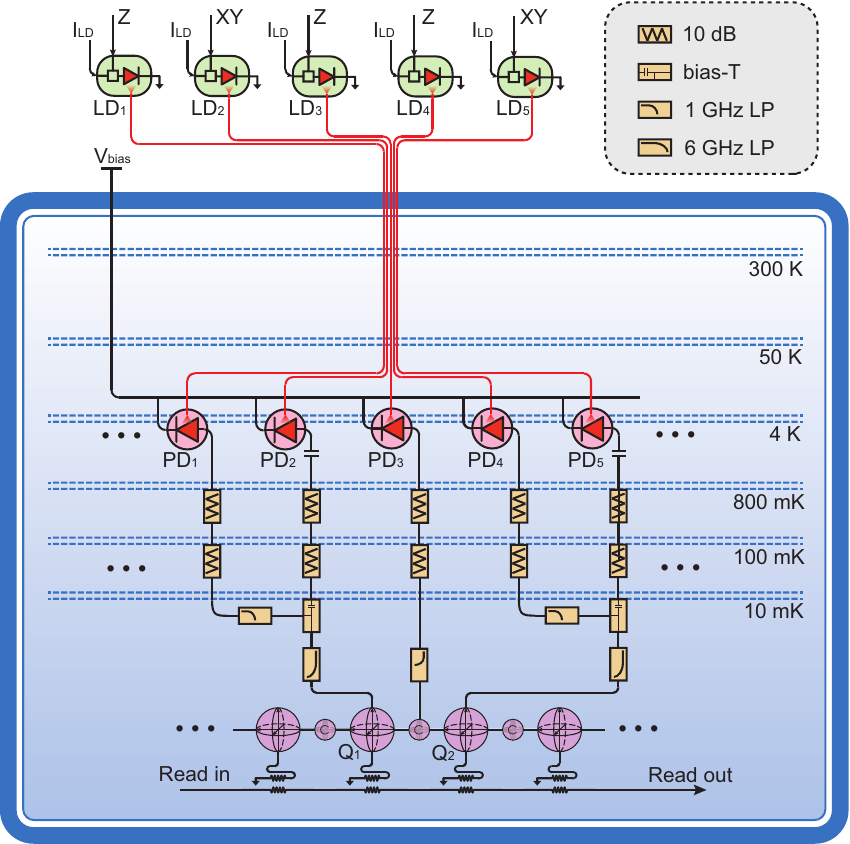}}
\caption{The setup for characterizing the noise and precision of the optical-assisted transmission line. InGaAs photodiodes (PDs) were installed at the 4 K stage of the DR with a reversed bias of 1 V. The coaxial cables for XY and Z signal between room temperature and 4 K were replaced by optical fiber. The IQ-mixed signal was amplified as the input of a laser diode (LD). The LD was a directly modulated laser with a center wavelength of 1490 nm.
}
\label{Fig:Setup}
\end{figure}

\begin{figure}[!tbp]
\centering
\resizebox{\columnwidth}{!}{\includegraphics{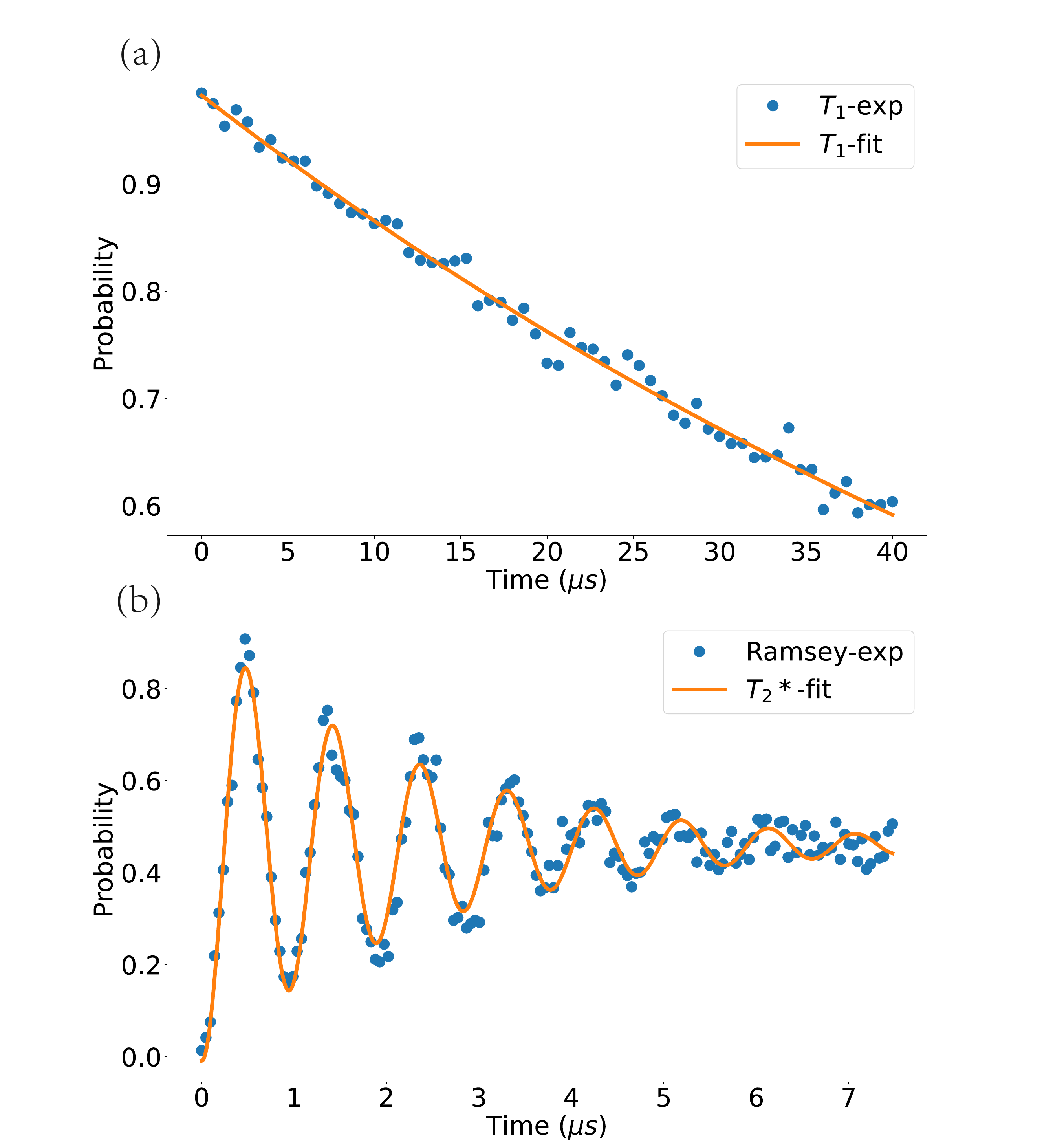}}
\caption{Characterizing the performance of the superconducting qubit with optically transmitted control signals. 
(a) and (b) A typical curve of $T_1$ and Ramsey measurement indicate a $T_1$ of 58.58 \(\pm\) 3.39 \(\mu\)s and a $T_2^*$ of 2.40 \(\pm\) 0.23 \(\mu\)s (409 MHz away from the sweet spot), respectively.}
\label{Fig:Result}
\end{figure}

The setup of controlling transmon superconducting qubits is shown in Fig. \ref{Fig:Setup}.
The control signals were produced in a homemade arbitrary waveform generator (AWG) \cite{10.1063/1.5120299, guo_control_2019}.
Specifically, the Z signals were square-wave pulses that were directly generated by the AWG.
The XY signals were microwave pulses with a frequency of several gigahertz and were up-converted in an IQ-Mixer.
Before being transmitted into the refrigerator, the signals were used to drive a directly modulated distributed feedback laser (DML) with a wavelength of 1490 nm \cite{li_optical_2024}.
Compared to externally modulating a laser with electro-optic modulators (EOMs) \cite{Lecocq2021}, DML exhibits better linearity that effectively suppresses higher harmonic while preventing the piezoelectric oscillation that distorts the square wave of Z pulse.
The modulated lasers are then transmitted into the DR by optical fibers.
InGaAs photodiodes are installed at the 4 K stage to convert the photons to photocurrent, thus regenerating the control signal for qubits.
InGaAs PDs with PIN structure offer a remarkable high-frequency response in the spectral region of NIR and thus have been widely used in high-speed optical communication with a data rate of tens of gigabits per second.
For analog signals, however, any minor distortion would introduce errors.
Thus, the electronic wiring is carefully designed to match the impedance well.

We employed two neighboring transmon qubits on a quantum processor to characterize the performance and noise characteristics of optically transmitted control signals.
The qubits were coupled to a shared readout cavity, enabling simultaneous dispersive measurement of their quantum states.
Additionally, the coupling strength between the two qubits was dynamically tuned using a tunable coupler, allowing precise control over their interaction.
The photocurrent noise with a directly modulated laser (DML) can be expressed as $S_I(\omega)=2eI+S_I^{\delta V}(\omega)+S_I^{RIN} (\omega)$, with the driving noise $S_I^{\delta V}(\omega)=S_V(\omega)I^2 s_{LD}^2/R_{LD}^2$ and the laser intensity noise $S_I^{RIN} (\omega)=I^2RIN(\omega)$.
Here, $s_{LD}$ and $R_{LD}$ are the slope efficiency (typically at the level of 0.2 W/A) and load impedance (normally 50 \(\Omega\)) of the DML, respectively.
$RIN(\omega)$ is the relative intensity noise.
A decrease in the photocurrent decreases noise but increases the required time for a quantum gate, thus increasing the decoherence.
Therefore, a tradeoff should be taken in selecting an appropriate photocurrent.
Considering both effects, we determine the photocurrent to be 103 \(\mu\)A for the single qubit gate. 

\section{Qubit characterization and noise analysis}

To characterize the decoherence performance of the qubit, we conducted measurements of energy relaxation $T_1$ and Ramsey experiments, respectively, as shown in Fig. \ref{Fig:Result}(a) and (b). 
$T_1$ and $T_2^*$ are measured to be 58.58 \(\pm\) 3.39 \(\mu\)s and 2.40 \(\pm\) 0.23 \(\mu\)s respectively.
Due to the qubit being detuned from the flux sweet spot by 409 MHz, the dephasing decay dominantly takes the form of $T_{\phi2}$.
By fitting it with $V(\tau) = A \cos(\delta \omega \tau + \phi_0)e^{-(\frac{\tau}{T_{\phi2}})^2 - \frac{\tau}{T_{\phi1}} }+B$, the corresponding $T_{\phi2}$ is obtained as 2.47 \(\pm\) 0.49 \(\mu\)s.
Here \(\tau\) denotes the evolution time, \(\delta\omega\) signifies the applied Ramsey oscillation frequency, while \( A \), \( B \), and \( \phi_0 \) are the fitting parameters. \( T_{\phi1} \) represents the time constant for white noise dephasing, and \( T_{\phi2} \) represents the time constant for correlated noise.
We can respectively estimate that the errors of the 50 ns idle gate caused by \( T_1 \) and \( T_{\phi 2} \) are 0.028\% and $0.013\%$~\cite{Malley2015qubit}. This serves as the basis for achieving high-fidelity quantum gates.

\begin{figure}[!tbp]\raggedright
\resizebox{\columnwidth}{!}{\includegraphics{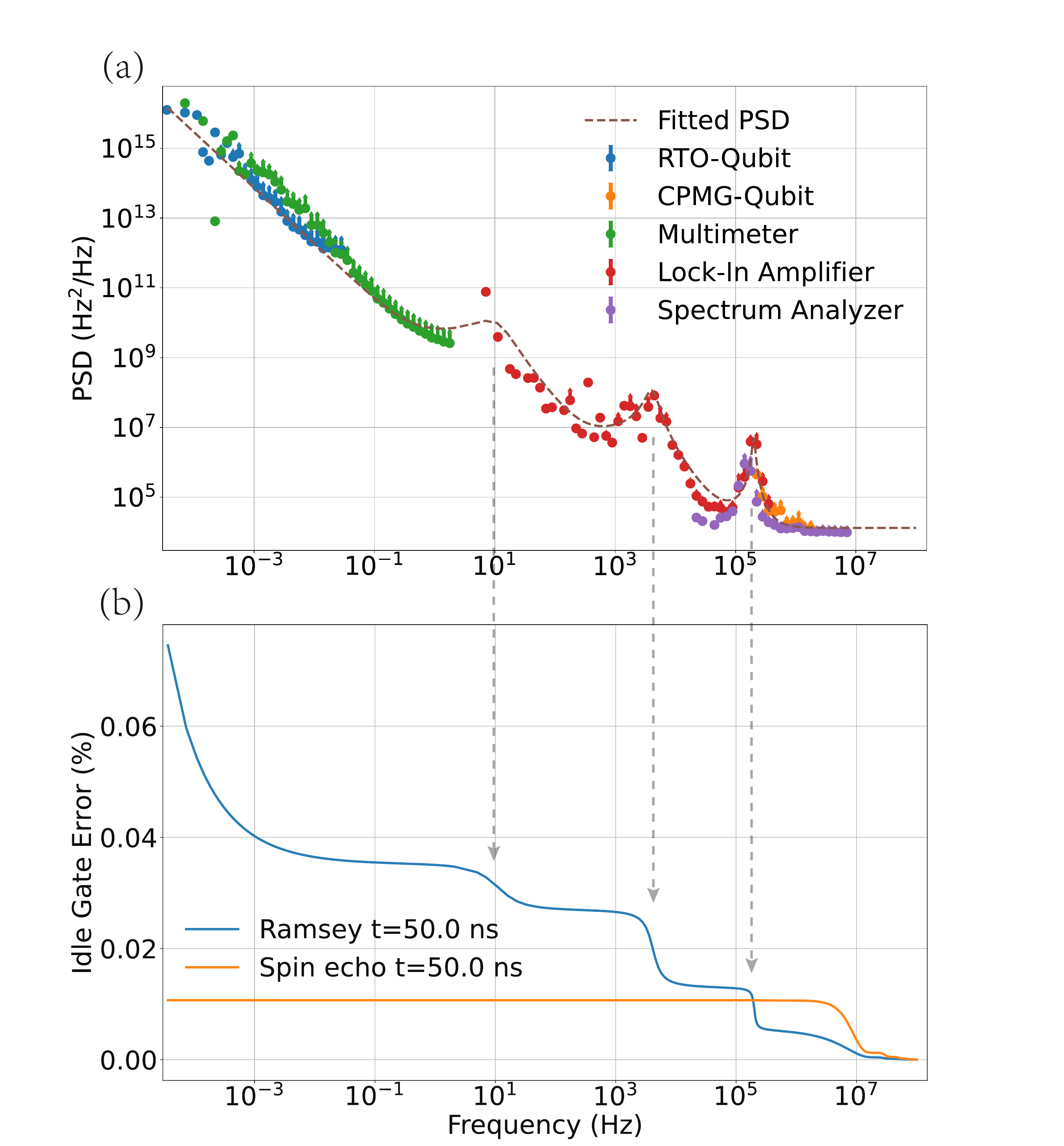}}
\caption{Characterizing the noise on optically transmitted Z control line. (a) Measured noise power spectrum densities (PSD) by a range of different methods, including the RTO and CPMG measurement on the qubit and several traditional methods at room temperature, covering the frequency from \(10^{-3}\) to \(10^7\) Hz. The dashed line represents the corresponding fitting result.  
(b) Based on the PSD results, the error rates of the Idle gates in the Ramsey and Spin echo scenarios are estimated, respectively. 
As shown by the dotted-line arrow in the figure, although several typical Lorentzian-type noise peak exists, the error rate integrated up to $10^{-3}$ Hz is still sufficiently small, indicating that the noise on the optical transmit control line is low enough.
}
\label{Fig:NoiseFig}
\end{figure}

For frequency-tunable transmon qubits, flux noise influences qubit dephasing time, highlighting the necessity of characterizing the noise power spectral density (PSD) on the Z control line.
We estimated the PSD of the optical transmitted Z control signal by the methods of Ramsey tomography oscilloscope (RTO) \cite{yan_spectroscopy_2012} (with time scale above \(\sim\) 1 s) and Carr, Purcell, Meiboom and Gill (CPMG) sequences  \cite{carr1954effects, meiboom1958modified, bylander2011noise} (with time scale from 500 ns to 25 \(\mu\)s) via the qubit.
The PSD is also verified through different measurements at room temperature, including the multimeter, lock-in amplifier, and spectrometer to cover a wide range of frequencies.
Detailed descriptions of these methods are provided in Appendix~\ref{app:noise}.
All the results of PSD measurement with different methods are illustrated in Fig. \ref{Fig:NoiseFig}(a).
The measured PSD of noise can be well fitted with the model of 1/f noise, white noise, and several Lorentzian noise,
\begin{equation}
    S_f(f) = S_0 + \frac{A}{f^{\alpha}} + \sum_{i=1}^{3} \frac{A_{L,i}}{1 + \left( \frac{f - f_{0,i}}{\Delta f_i} \right)^2}
    \label{Eq:s_f}
\end{equation},
shown as the dashed line in Fig. \ref{Fig:NoiseFig}(a).
We can then integrate the PSD with the weighting functions of Ramsey and Spin echo to obtain the root mean square of the phase noise and, further, the error rate of the Idle gate for different time scales, as shown in Fig. \ref{Fig:NoiseFig}(b).
As a reference, we notice that the Idle gate error rate corresponding to \(10^{-3} Hz\) is about 0.04\%, indicating a sufficiently low noise level of the optically transmitted Z control line.

\section{Quantum-gate benchmarking}

The optimization of single- and two-qubit gates is similar to the traditional configuration.
For the single-qubit X/2 gate, its frequency, amplitude, and Drag Alpha factor are iteratively calibrated. The frequency is calibrated through the Ramsey oscillation, whereas the amplitude and Drag Alpha factor are calibrated by measuring the fidelity of the transition from the $|0\rangle$ state to the $|1\rangle$ state with a large odd number of X gates.
For two-qubit CZ gate \cite{li_realisation_2019-1}, after roughly determining the length and amplitude to obtain the conditional phase and minimizing the state leakage, We amplify state leakage errors and conditional phase errors more accurately by superimposing multiple CZ gates, thereby obtaining more precise amplitude.
It is worth noting that the inevitable impedance mismatch in the circuit leads to a distortion of the square wave signal that should be corrected.
The corrected control signal could be presented as $V_c(t)=V_o(t)+\sum{V_i(t)}$ \cite{doi:10.1126/science.aaw1611}, where $V_i(t)$ is the i-th order correction voltage. The lowest (highest) voltage for the corrected signal $V_C(t)$ is usually lower (higher) than the original one $V_o(t)$ near the rising and falling edges.
Since the laser intensity, as well as the corresponding photocurrent, can only be a non-negative value, it is necessary to apply an additional bias for the correction.
Such bias is required throughout the lifetime of the qubit and can be terminated between trials to reduce the active heat load.

\begin{figure}[!tbp]\center
\resizebox{\columnwidth}{!}{\includegraphics{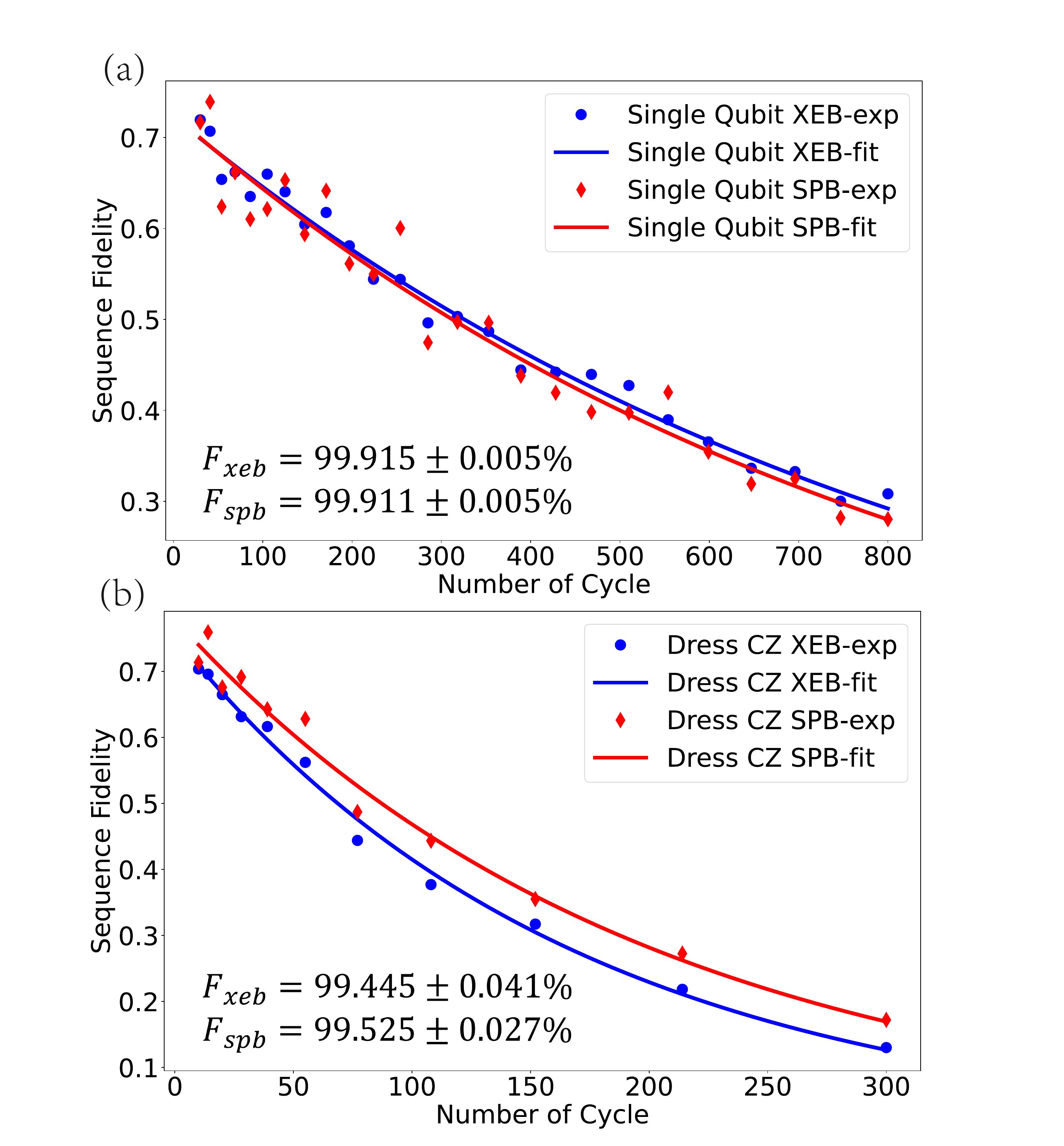}}
\caption{Speckle Purity Benchmarking (SPB) and Cross-Entropy Benchmarking (XEB) for single-qubit gates and two-qubit CZ gate. (a) The fidelity of a single qubit gate is 99.915\% \(\pm\) 0.005\%, with an average control error of 0.004\%. (b) The dressed fidelity of the two-qubit CZ gate is 99.445\% \(\pm\) 0.041\%, corresponding to a pure fidelity of 99.676\% \(\pm\) 0.041\% with an average control error of 0.053\%.}
\label{Fig:xeb and spb}
\end{figure}

Finally, the cross-entropy benchmarking (XEB) and Speckle Purity Benchmarking (SPB) \cite{Neill2018,Boixo2018} were performed to assess the fidelity and purity of single-qubit and two-qubit operations.
The control error can be estimated as the difference between them.
With a length of 50 ns for single-qubit X/2 gate, the average control error is estimated to be 0.004\% for the first qubit ($Q_1$) and 0.023\% for the second qubit ($Q_2$), and the average control error for a two-qubit CZ gate with length $38~ns$ between $Q_1$ and $Q_2$ is 0.053\% as shown in Fig. \ref{Fig:xeb and spb}.
Here, the control errors for both single- and two-qubit operations are far below the errors introduced by the decoherence of the qubits.
The pauli fidelity of single-qubit gates is estimated as 99.915\% \(\pm\) 0.005\% and 99.854\% \(\pm\) 0.014\% for $Q_1$ and $Q_2$ respectively.
The Pauli fidelity of the two-qubit CZ gate is 99.676\% \(\pm\) 0.041\%.
The estimated fidelity for both types of quantum gates has exceeded the requirement of surface code for realizing an error-correction-protected logic qubit \cite{PhysRevA.86.032324}.

\section{Discussion and conclusion}

In this work, we experimentally demonstrated an optical-assisted transmission line of control signals for superconducting qubits. The microwave control signal was modulated on the intensity of the laser by a direct modulated LD at room temperature. The laser is transmitted into the refrigerator and regenerates the microwave control signal at the 4 K plate without the need for coaxial cables. Therefore, the heat load of DR was significantly loosened from room temperature to the 4 K plate, benefiting from the low thermal conductivity of optical fiber. The thermal dissipation was carefully balanced so that commercial DR could support thousands of channels.
Except for the power supply, the optical-assisted transmission line has two microwave interfaces for signal input and output, respectively. Thus, it acts just like a regular RF cable in the system, which makes it downward compatible with the existing electronics control system. It is able to ``plug-and-play'' without making any changes to the current control system.
Furthermore, with coaxial cables replaced by optical fibers, it is much more flexible to configure the control system. For instance, the DR can be far away from the control system to reduce noise and vibration. Particularly, with the use of hollow core fiber \cite{Poletti2013}, the transmission of the signal is close to the speed of light in the vacuum, thus considerably reducing the time cost of feedback control of qubits.

\begin{acknowledgments}
We acknowledge insightful discussions with Sheng-Kai Liao, Yuan Cao, and Juan Yin.
This work was supported by Cultivation Project of Shanghai Research Center for Quantum Sciences (Grant No. LZPY2024), the Innovation Program for Quantum Science and Technology (Grant No. 2021ZD0300200), Shanghai Municipal Science and Technology Major Project (Grant No.2019SHZDZX01), the National Natural Science Foundation of China (Grant No. 12174374, No. 12274464), Shanghai Rising-Star Program (Grant No. 21QA1409600, 23QA1410000), the Shanghai Sailing Program (Grant No. 21YF1452500), CAS Young Interdisciplinary Innovation Team (Grants No. JCTD-2022-20), Anhui Initiative in Quantum Information Technologies, China Postdoctoral Science Foundation, and the Chinese Academy of Sciences.
H.-L. Huang acknowledges support from the Natural Science Foundation of Henan (Grant No. 242300421049).
Y.-H. Li and M. Gong were supported by the Youth Innovation Promotion Association of CAS (Grant No. 2023475, No. 2022460).
\end{acknowledgments}

\section*{Data Availability}
The data that support the findings of this article are available from the corresponding author upon reasonable request.

\appendix

\setcounter{figure}{0}
\setcounter{table}{0}
\renewcommand{\thefigure}{A\arabic{figure}}
\renewcommand{\thetable}{A\Roman{table}}

\section{Cryogenic photodiode response and stability}
\label{app:pd}
\subsection{Low-temperature response}

\begin{figure}[!tbp]
    \centering
\includegraphics[width=1.0\linewidth]{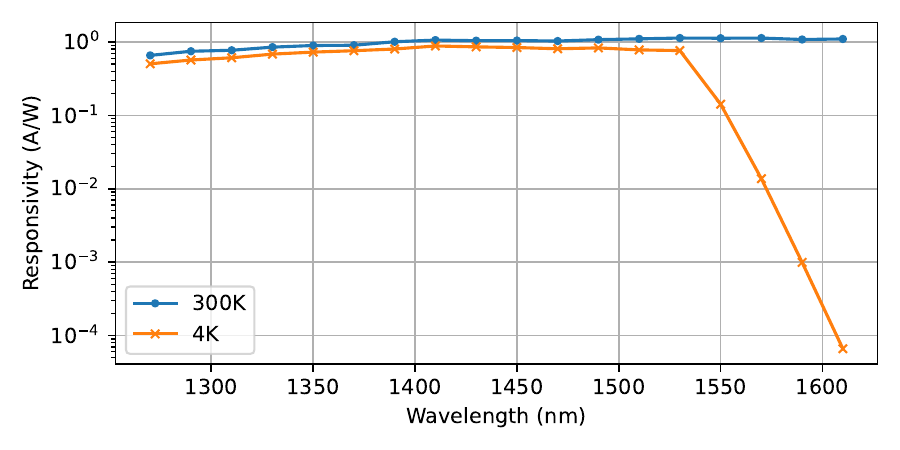}
    \caption{The responsivity of photodiode under 300 K and 4 K.}
    \label{fig:resp}
\end{figure}

InGaAs PDs with PIN structure offer a remarkable high-frequency response in the spectral region of NIR and thus have been widely used in high-speed optical communication with a data rate of tens of gigabits per second.
For the photoelectric effect to occur, the energy of a single photon should exceed the energy gap $E_g$ of the semiconductor materials of the PD to emit a free electron.
InGaAs PD has a bandgap of InGaAs of 0.735 eV \cite{10.1117/12.897576} at room temperature, corresponding to a cutoff wavelength of 1687 nm.
With the decrease of temperature, the value of bandgap increases following $E_g(T)=E_g(0)-\sigma T$ \cite{10.1117/12.897576, BANDARA2007211}, where $E_g(0)=$~0.822 eV is the bandgap of InGaAs at 0 K, $\sigma=3\times 10^{-4}$~eV/K, and $T$ is the temperature.
That is, the cutoff wavelength reduces to 1508 \(\sim\) 1511 nm when the PD is installed in the DR with a temperature between 10 mK to 4 K.
More specifically, $E_g(0K) = 0.822$ eV (1508 nm according to $E_g=hc/\lambda$, $h$ is the Planck constant and $c$ is the speed of light in vacuum), $E_g(1K) = 0.8217$ eV (1509 nm), $E_g(4K) = 0.8208$ eV (1511 nm), and $E_g(77K) = 0.7989$ eV (1552 nm).
The cutoff wavelength at 4 K is verified in our setup, as shown in Fig. \ref{fig:resp}.
On the other hand, since at most one free electron can be emitted by absorbing one photon according to the photoelectric effect, a longer wavelength would be beneficial to reduce the heat load.
Thus, the wavelength of 1490 nm was chosen so that almost all optical elements designed for the C-band also work well at this wavelength.
It is worth noting that components like photomultiplier or avalanche PD can efficiently generate multiple electrons with one photon.
However, it requires a high voltage of bias that introduces unacceptable high heat dissipation; thus, it is not taken into account here.


\subsection{Polarization dependence and long-term stability}

\begin{figure}[!tbp]\center
\resizebox{\columnwidth}{!}{\includegraphics{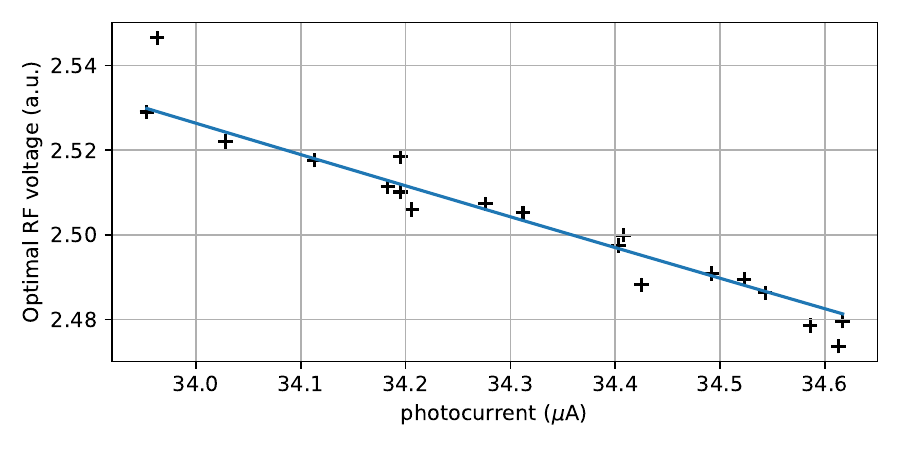}}
\caption{(color online). The relationship between the optimal control voltage of the single qubit $X$ gate and the photocurrent with the polarization of incident laser changes. The control voltage is optimized by fitting the result of Rabi oscillation.}
\label{Fig:Stability}
\end{figure}

One of the potential sources of instability is the change of polarization in fiber.
The response of PD is slightly dependent on the polarization of the injected laser, owing to the anisotropy of materials and the imperfection of assembling.
Suppose the polarization of the laser shine on the surface of PD is $\rho=p|H\rangle\langle H|+(1-p)|L\rangle\langle L|$, where $|H\rangle$ ($|L\rangle$) is the polarization that the PD has highest (lowest) response, $0\le p\le1$.
The overall response of the PD is represented as $R=pR_H+(1-p)R_H(1-D)$, where $R_H$ is the response with input state $|H\rangle$, $D$ is the polarization dependence loss.
Thus, when the polarization of the laser changes, the maximal fluctuation of the response is $\Delta_R={R_{max}}/{R_{min}}-1=D/(1-D)$.
Normally, $D$ is at the level of $10^{-2}\sim10^{-3}$.
By disturbing the fiber through a polarization controller, the response photocurrent of the PD varies, and the optimal driven power on the LD to product $X$ gate changes accordingly.
The relationship between these two values is shown in Fig. \ref{Fig:Stability}.
Thus, it is important to keep the polarization of the laser stable and prevent the temperature drift and stress variation of the fiber to obtain long-term stability.

\section{Thermal-load calculations}
\label{app:thermal}
\subsection{Photodiode active heat load}

As the passive heat load is negligible, the active heat load is the main issue to be considered in our configuration.
When regenerating the control signal by photocurrent, the additional heat dissipation can be estimated as
\begin{equation}
\label{Eq:P_LOAD}
P_{heat} = (V_{bias} + \frac{h\nu}{e\eta})\sqrt{\frac{\kappa P_{gate}}{Z}} - P_{gate}.
\end{equation}
Here, $V_{bias}$ is the reverse bias of the PD for increasing the response at high frequency, which is about \(1V\) in our setup.
$Z$ stands for the load impedance, which is normally 50 $\Omega$.
$e$ stands for the elementary charge, $h$ represent the Planck constant, and $\nu$ is the frequency of photon.
$\eta$ is the conversion efficiency, including the effect of alignment accuracy and quantum efficiency.
Generally speaking, a lower reverse bias on PD and longer wavelength of photon result in lower heat dissipation.
$P_{gate}$ is the required power of the expected quantum gate.
For microwave gate, $\kappa = 2$, while for square wave gate, $\kappa = 1$. 
$P_{gate}$ is highly dependent on the design of the qubit and the cryogenic wiring.
For a rough estimation, we assume $P_{gate}=-66$ dBm for XY gate, and $P_{gate}=-27$ dBm (corresponding to a current of 200 \(\mu\)A).

For the calculation of heat load, it can mainly be divided into passive heat load and active heat dissipation. The passive heat load is mainly caused by the heat flow from high temperature to low temperature, which is related to the thermal conductivity, cross-sectional area and length of the transmission line, and can be estimated by the following formula:
\begin{align}
    P = \int_{T_1}^{T_2} \frac{\sum{\rho_i(T) A_i}}{L} dT
\end{align}
where $\rho_i$ and $A_i$ are the thermal conductivity and cross-section respectively, and $L$ is the length of the transmission line. 

For traditional coaxial cables made of currently commonly-used materials, such as niobium-titanium cables and stainless-steel cables, their passive heat load at the 4K layer is approximately 1 mW \cite{Krinner2019}. Calculated based on a refrigeration power of 1.5 W, the upper limit is about 1500 cables. Correspondingly, due to the extremely low thermal conductivity of optical fiber materials, the passive heat load calculated at the 4K layer is only about 5.6 $\mu$W \cite{joshi2023scaling}. This means that the number of optical fiber links that can be accommodated is two orders of magnitude higher than that of traditional coaxial cables.

\subsection{Active heat dissipation}
Regarding active heat dissipation, it can mainly be divided into the attenuation caused by attenuators and line losses, as well as the heat dissipation of active devices, such as the photodiode adopted in this paper. The heat dissipation caused by attenuators and losses is the attenuated energy. It mainly depends on the configuration of attenuators at each stage of the dilution refrigerator and the energy of the signal before attenuation. The configuration of attenuators is related to the signal-to-noise ratio (SNR) of the original signal, the temperature at each stage of the dilution refrigerator, and the mutual inductance of the control lines. Heat dissipation can be effectively reduced by improving the SNR and reducing the mutual inductance intensity of the control lines.

\begin{figure*}[tb]\centering
\resizebox{\textwidth}{!}{\includegraphics{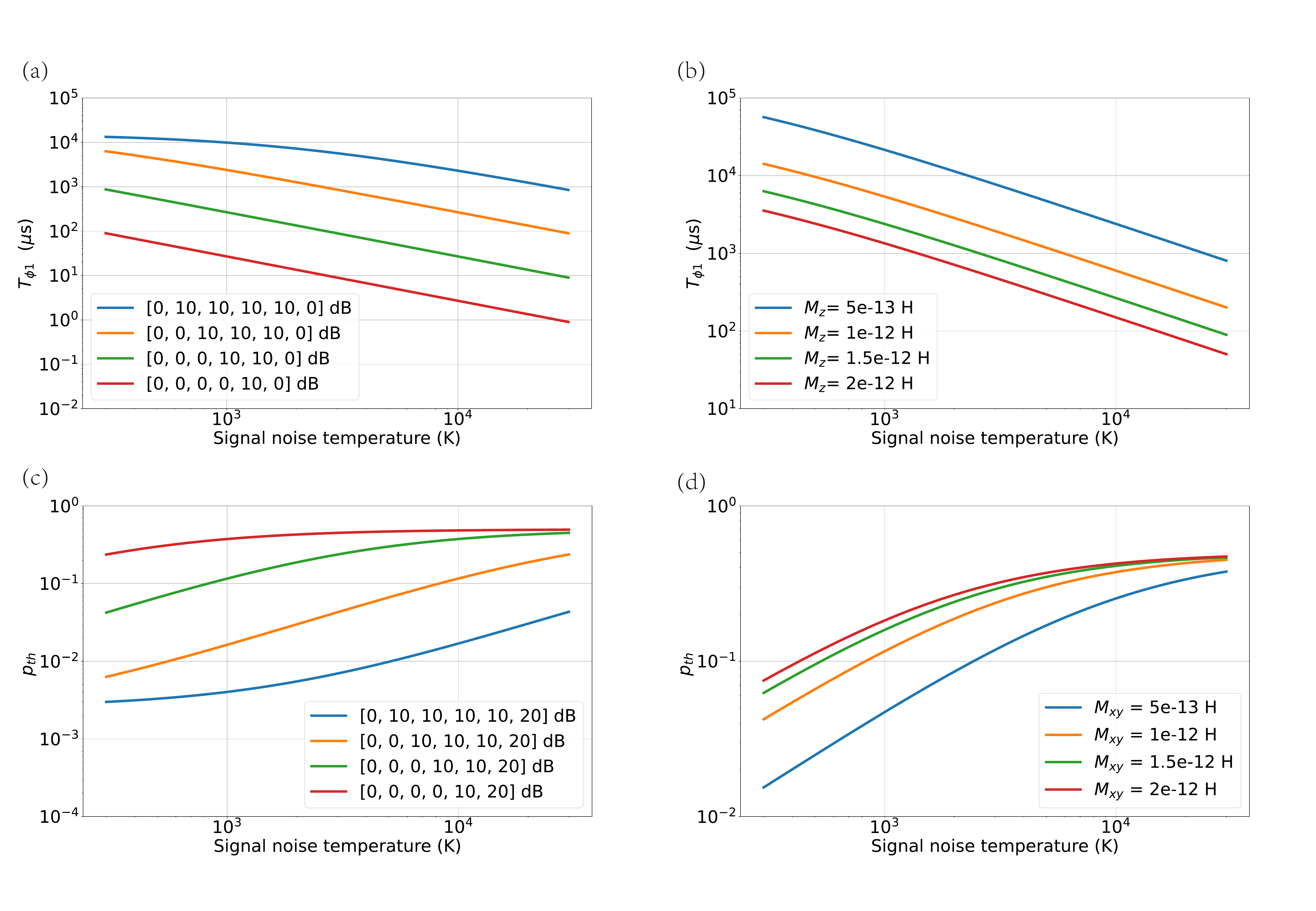}}
\caption{The influence of signal noise on qubits. (a) Influence of the noise temperature of the Z-control signal (generated at room temperature) on the qubit's $T_{\phi1}$ under varying attenuators. The simulation results in the figure assume a qubit sweet frequency of 5.3\,GHz, an idle frequency of 5\,GHz, and a mutual inductance of 1.5\,pH for the Z-control signal. The attenuator values in the figure correspond to the dilution refrigerator stages [RT, 50K, 4K, Still, CP, MXC]. (b) Similar to (a), this shows the influence of different mutual inductance strengths on the qubit's $T_{\phi1}$ with a fixed attenuator configuration of $[0, 0, 10, 10, 10, 0]$. (c) Influence of the noise temperature of the XY-control signal (generated at room temperature) on the qubit's thermal excitation rate under varying attenuators. The mutual inductance of the XY-control signal used in the simulation results of the figure is 1\,pH. (d)  Similar to (c), this shows the influence of different mutual inductance strengths on the qubit's thermal excitation rate with a fixed attenuator configuration of $[0, 0, 0, 10, 10, 20]$.}
\label{Fig:heat load}
\end{figure*}

To describe this process more clearly, Fig. \ref{Fig:heat load}  considers the influences of the signal's noise temperature, the attenuators in the transmission link (wherein transmission line losses are not additionally considered and can also be simplified as an attenuator model), and the mutual inductance between the control line and the qubit on the qubit, respectively. Figs. \ref{Fig:heat load} (a) and (b) respectively show the influence of the Z-control signal on the qubit's $T_{\phi1}$ under different attenuator configurations and mutual inductance strengths. Currently, the $T_{\phi1}$ of quantum processors with tunable qubit frequencies has reached the order of 100 $\mu s$ ~\cite{gao2025establishing,google2025quantum}.
Figs. \ref{Fig:heat load} (c) and (d) consider the influence of the XY-control signal on the qubit's thermal excitation rate, which determines the upper limit of readout fidelity and typically needs to be less than $1\%$. 
Currently, there are also many initial state reset schemes to mitigate the impact of thermal excitation ~\cite{riste2012initialization,marques2023all,gao2025establishing}.
As can be seen from the Figs. \ref{Fig:heat load}, a greater number of attenuators can significantly improve qubit performance; however, this also poses challenges to the cooling power of the dilution refrigerator. Table \ref{tab:xy dissipation} and Table \ref{tab:z dissipation} calculate the active heat dissipation induced by attenuator configurations for the XY and Z control signals, respectively, in the traditional coaxial cable scheme, with typical values as examples.
For the active heat dissipation of XY signals, it can be observed that it is primarily limited by the CP and MXC layers. On one hand, the attenuation intensity can be reduced by further decreasing the mutual inductance of XY signals. On the other hand, since XY signals are only applied during gate operations, active heat dissipation can be further mitigated by decreasing the duty cycle.
For the active heat dissipation of Z control signals, it can be seen that adding attenuators in the CP layer is actually unwise. Considering the SNR, it is entirely feasible to place the attenuators in higher-temperature layers. Let's consider a relatively ideal condition where the signal SNR is sufficiently good and the noise temperature at the room temperature end is 300K. Even if we only add 10dB attenuation in the 4K layer and reduce the mutual inductance of the Z control lines to 0.4pH, the corresponding Tphi1 can still reach over 1ms, thereby solving the problem of insufficient cooling power in the CP layer and even the Still layer. 
In summary, active heat dissipation can definitely be optimized.

\begin{table*}[htbp]
  \centering
  \caption{The heat dissipation generated by the XY control signal in different temperature layers of the dilution refrigerator.}
  \begin{tabular}{|c|c|c|c|c|c|}
    \hline
    Stage & $P_C$ (W) & XY attenuators(dB) & XY power (dBm) & $P_L$ (W) & $P_C$/$P_L$ \\ \hline
    50K   & 30      & 10                 & -16            & $2.261 \times 10^{-4}$ & 132702.4 \\ \hline
    4K    & 1.5     & 10                 & -26            & $2.261 \times 10^{-5}$ & 66351.2  \\ \hline
    Still & 0.04    & 10                 & -36            & $2.261 \times 10^{-6}$ & 17693.65 \\ \hline
    CP    & $2 \times 10^{-4}$ & 10                 & -46            & $2.261 \times 10^{-7}$ & 884.683  \\ \hline
    MXC   & $1.9 \times 10^{-5}$ & 20                 & -66            & $2.487 \times 10^{-8}$ & 764.044  \\ \hline
  \end{tabular}
  \label{tab:xy dissipation}
\end{table*}

\begin{table*}[htbp]
    \centering
    \caption{The heat dissipation generated by the Z control signal in different temperature layers of the dilution refrigerator.}
    \begin{tabular}{|c|c|c|c|c|c|}
        \hline
        Stage & $P_C$ (W) & Z attenuation (dB) & Z current (A) & $P_L$ (W) &  $P_C$/$P_L$ \\ \hline
        50K   & $30$ & 10            & $6.320\times10^{-3}$ & $3.601\times10^{-3}$  & $8.332\times10^{3}$  \\ \hline
        4K    & $1.5$ & 10            & $2.000\times10^{-3}$ & $3.594\times10^{-4}$  & $4.173\times10^{3}$  \\ \hline
        Still & 0.04 & 10   & $6.325\times10^{-4}$ & $3.600\times10^{-5}$  & $1.111\times10^{3}$  \\ \hline
        CP    & $2 \times10^{-4}$ & 10            & $2.000\times10^{-4}$ & $3.600\times10^{-6}$  & $5.556\times10^{1}$  \\ \hline
        MXC   & $1.9 \times10^{-5}$ & 0             & $2.000\times10^{-4}$ & 0  &   \textbackslash{} \\ \hline
    \end{tabular}
    \label{tab:z dissipation}
\end{table*}

\section{Noise-spectrum measurement and gate-error estimation}
\label{app:noise}
\subsection{Room-temperature instrumentation}
Here, we will introduce how to test the PSD of the entire frequency band of the Z control signal by using electronic devices at room temperature. 

First, we conduct the test of the noise spectral density at extremely low frequencies.
By connecting the photocurrent generated at room temperature to a multimeter and collecting the photocurrent signal over an extended period, a time-varying photocurrent signal can be obtained. Performing a fast Fourier transform(FFT) on the photocurrent in the time domain can yield the PSD in the frequency domain. 
The longer the collection time, the lower the corresponding spectral frequency and the shorter the frequency interval. 
On the other hand, the shorter the collection time interval, the higher the corresponding maximum frequency.

Secondly, we are able to utilize a lock-in amplifier to measure the PSD in the frequency range from 3 Hz to 500 kHz.
The operating principle of a lock-in amplifier involves feeding the signal under measurement and the reference signal into a mixer. This process generates frequency information related to the difference and sum of the frequencies of these two signals.
When the frequency $f$ of the signal being measured matches that of the reference signal, the frequency-difference component is transformed into a DC signal.
The signal emerging from the mixer then passes through a low pass filter, after which the frequency-converted signal is collected. 
The amplitude of the measured frequency-converted DC signal corresponds to the amplitude of the signal with a frequency of $f$.
By adjusting the frequency of the reference signal, we can obtain the PSD of the signal under measurement at various frequencies. 

The above two methods are mainly used to characterize the relatively low-frequency noise. 
In addition, a spectrum analyzer can be used to test the noise spectrum in the kHz-MHz frequency band.
The principle of the spectrum analyzer for noise spectrum testing in the low-frequency band is basically the same as that of the multimeter. 
It just uses a data acquisition card with a higher sampling rate to collect the time-domain signals and performs frequency-domain characterization through FFT.
At high frequencies, its principle is similar to that of the lock-in amplifier. 
It uses a local oscillator microwave signal to convert the frequency of the measured signal to the center frequency of the subsequent band-pass filter and collects the filtered signal to obtain the amplitude information of the corresponding frequency.

\subsection{Qubit-based measurements}
Here, we estimate the noise power spectral density (PSD) in the frequency ranges from \(10^{-4}~\text{Hz}\) to \(10^{-1}~\text{Hz}\) and from \(10^{4}~\text{Hz}\) to \(10^{6}~\text{Hz}\) through qubit.
The PSD of the frequency $S_f (f)$ was estimated by the method of Ramsey tomography oscilloscope (RTO) \cite{yan_spectroscopy_2012} (with time scale above $\sim$ 1 s) and Carr, Purcell, Meiboom and Gill (CPMG) sequences  \cite{carr1954effects, meiboom1958modified, bylander2011noise} (with time scale from 500 ns to 25 \(\mu\)s).
With RTO method, the qubit was initially prepared to $|\Psi\rangle=\frac{1}{\sqrt{2}}(|0\rangle + |1\rangle)$ and measured in the $\frac{1}{\sqrt{2}}(|0\rangle \pm |1\rangle)$ basis after a certain time duration $\tau$ to estimate the accumulated relative phase $\delta \phi(t)$ between $|0\rangle$ and $|1\rangle$.
When only considering noise with a frequency far lower than the inverse of the qubit lifetime, $\delta \phi (t)=2\pi\tau\delta f_{01}(t)$.
Through repeated trials, a series of $\delta f_{01}(t)$ could be estimated by measuring $\delta \phi(t)$, and then the PSD can be obtained by performing an FFT on $\delta f_{01}(t)$.
With the CPMG method, $S_f (f)$ was obtained by incorporating additional X-gates into the spin echo circuit.
This is akin to the effect of a band-pass filter, where the center frequency of this filter depends on the decay time and the number of X-gates applied.
Therefore, by varying these two parameters, one can obtain PSD at different frequencies.

\subsection{Conversion between qubit-frequency noise and voltage noise}
The qubit frequency $f_{01}$ and the current of the Z control signal at 4K stage satisfy:
\begin{equation}
    f_{01} = (f_{\text{01max}}-f_{\text{ah}}) \sqrt{cos[\pi k_f (I-I_{\text{offset}})]}+f_{\text{ah}},
\end{equation}
where $f_{\text{01max}}$ is the maximum frequency, $f_{\text{ah}}$ is anharmonicity of the qubit, $k_f$ is the coefficient of quantum flux and current, and $I_{\text{offset}}$ is local bias.
Therefore, the relationship between the PSD of the qubit's frequency and that of the control signal's current is related according to 
\begin{equation}
    S_V(f)=\left(50\frac{dI}{df}\right)^2 S_f(f). 
\end{equation}
Thus, we can unify the units of the PSDs measured at room temperature and low temperature.

\subsection{Idle-gate error estimation}
For the given power spectral density \(S_f(f)\) of the frequency noise of a qubit, the corresponding root mean square of the phase noise can be obtained by the following equation: 
\begin{align}
    \left\langle \phi_n^2(t) \right\rangle 
    &= \int_{0}^{\infty} df S_f(f) W(f) 
    \label{Eq:RMS}
\end{align}
where \(W(f)\) represents the spectral weight function. 
For the Ramsey and spin echo experiments, their weighting functions are as follows:
\begin{align}
    W_{\text{R}}(f) &= \frac{\sin^2(\pi f t)}{(\pi f)^2} \\
    W_{\text{SE}}(f)&= \tan^2 \left( \frac{\pi f t}{2} \right) \frac{\sin^2(\pi f t)}{(\pi f)^2}.
\end{align}
Thus, the error rate of the corresponding Idle gate can be estimated by the following equation~\cite{Malley2015qubit}:
\begin{align}
    e_{I,\phi} = \frac{1}{6} \left\langle \phi_n^2(t) \right\rangle
\end{align}
Influenced by the weighting function, the upper limit of the integral in Eq.\ref{Eq:RMS} need only be taken up to 100 MHz. Meanwhile, the lower limit of the integral reflects how the gate error rate behaves on a long-time scale.

\bibliographystyle{apsrev4-2}
\bibliography{SCO}

\end{document}